\documentclass[aps,twocolumn]{revtex4}

\usepackage{graphics}

\begin{document}
\voffset 1truein

\bibliographystyle{apsrev}

\title{Missing Lorenz-boosted Circles-in-the-sky}

\author{Janna Levin}
\affiliation{Barnard College of Columbia University,
Department of Physics and Astronomy,
3009 Broadway, NY, NY 10027}
%\affiliation{Astrophysics, Oxford University, Denys Wilkinson Building, Oxford OX1 3RH}

\begin{abstract}
A topologically finite universe, smaller than the observable horizon, will have
circles-in-the-sky: pairs of circles around which the temperature
fluctuations in the cosmic microwave background 
are correlated. The circles occur along the intersection 
of copies of the spherical surface of last scattering. For any
observer moving with respect to the microwave background, the circles
will be deformed into ovals. The ovals will also be displaced relative
to the direction they appear in a comoving frame. The
displacement is
the larger of the two effects.
%The deviation of the angular radius of the oval from a perfect circle 
In a Lorenz boosted frame, the angular displacement of a point on 
the surface of last scattering relative to the comoving frame 
is proportional
to the velocity. For the Earth's motion, the effect is on the order of 
$0.14^o$
at the very worst. If we live in a small universe and are
looking for an identical copy of a spot in the sky, it may be
displaced by as much as $0.14^o$ from where we expect. 
This can
affect all pattern based searches for the topology of the universe.
In particular, high-resolution
searches for circle pairs could be off by this much.

\end{abstract}

\maketitle

The earth is nearly comoving with the expansion of the
universe but not quite. The entire galaxy moves with respect to the 
cosmic microwave 
background (CMB) at a speed $\beta=v/c = 1.23 \times 10^{-3}$
relative to the speed of light. The motion of the earth 
creates a large dipole-fluctuation in the CMB temperature since the
universe looks hotter in the direction of our motion than in the
opposite direction. The dipole is subtracted from data sets such as
those from COBE \cite{cobe} and WMAP \cite{wmap} to remove
noncosmological contributions to the maps. Subtracting the dipole
alone does not correct the map for any distortions in the shape or
location of features in the sky. Ordinarily this doesn't matter in the
least since the universe is assumed to be homogeneous and
isotropic. Analysts are interested in angular averages over the sky,
not the precise location of a given hot or cold spot.

In a finite universe, homogeneity and isotropy cannot be assumed
globally since topological identifications nearly always break these
symmetries (see reviews \cite{{wolf},{ellis},{review1},{review2}}). A
search of the CMB 
data for evidence of a finite universe relies on detailed information
on the shape and location of features in the sky. These pattern-based
searches \cite{{css},{spots},{deol}} all ignore the motion of the
earth with respect to the CMB. As shown below, there is a small
effect, yet none-the-less relevant to pattern-based searches, due to
the earth's motion.

In a topologically compact space, there are a preferred
set of observers for whom the volume of space is smallest
\cite{{bl1},{bl2}}. In the preferred frame, 
space is topologically identified 
but there is no mixing of the time component in 
the identification rules. This frame naturally coincides with the 
comoving frame and the preferred observers are at rest with respect to 
the expansion of the universe \cite{bl2}.
For the sake of argument,
consider one such hypothetical observer, C, completely at rest 
with respect to the CMB and living in a flat spacetime.
Working in his comoving frame,
last scattering occurred at a time $\eta_0 - \eta_*$ in the
past where
$\eta_0$ is the age of the universe today and $\eta_*$ is the 
time of last scattering. For the sake of argument we will take last
scattering
to happen at an instant. Since 
our experiments measuring the microwave background
are
effectively instantaneous relative to the age of the universe, all of
the 
light collected traveled exactly the same distance in all directions
thereby defining a spherical surface of radius $\eta_0-\eta_*=r$. 
All the CMB light C observes originated on this
surface of last scattering.

\begin{figure}[ht]
\vspace{40mm}\includegraphics{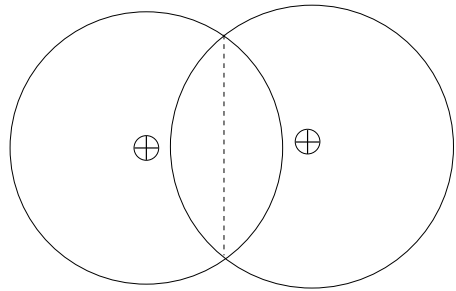} 
\vskip 5truept
\caption{The surface of last scattering intersects with itself in 
a finite universe smaller than the diameter of the surface. The
self-intersection can most easily be visualized in a tiling picture,
such as the above where a two-dimensional slice through space is
shown. The intersection of the two spheres occurs along the dotted-line
circle. When an observer at rest with respect to the expansion looks
to the right, he sees temperature fluctuations along the
dotted-line. When he looks to the left he sees identical temperature
fluctuations along the dotted-line. Therefore he measures a correlated
pair
of circles, one to the right, the other to the left.
}
\label{circles}
\end{figure}

\begin{figure}[ht]
\vspace{40mm}\includegraphics{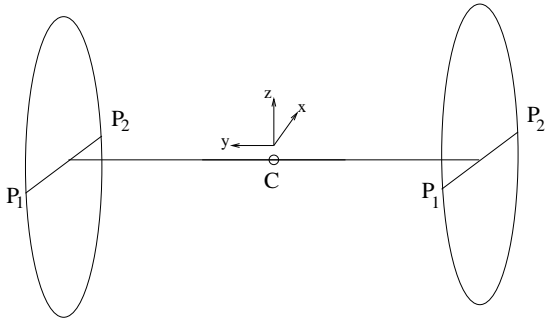} 
\vskip 5truept
\caption{Another view of the correlated circle pair of figure
\protect{\ref{circles}}. C sees one circle in front of him and one
behind him.
In general, circle pairs may be out of
  phase and not face-to-face.
}
\label{onecirc}
\end{figure}

A finite universe looks like an infinite universe tiled with an infinite number
of copies of the fundamental space. In each tile is an identical copy of the
earth and each image of the earth is encapsulated by an image of
the surface of last 
scattering. Some of these copies of the surface of last scattering will be
near enough to each other in the tiling to overlap. The overlap occurs along
a circle so that observer C will see identical variations in the
temperature fluctuations occurring along circle pairs \cite{css}
as illustrated in figures \ref{circles} and \ref{onecirc}. These
circles-in-the-sky, as they were coined when found in Ref.\
\cite{css}, are particularly 
important observationally since they occur in any topologically compact space
and in principle can be detected without any prior assumption about the 
shape and size of space, as long as the universe is smaller than the 
surface of last scattering.

In principle, a discovery of circles-in-the-sky would be an unambiguous
and definitive observation of topology. In practice, there are impediments
to making such an observation. Lensing of photons along the line of sight
from decoupling until today can deflect fluctuations off the circle or 
damage the correlation between circle pairs, as can reionization or an
integrated Sachs-Wolfe contribution. Other important effects
include
a finite thickness to the surface of last scattering as well as
Doppler
effects on small scales.
Matter oscillations can Doppler shift the CMB in a directionally
dependent way.(In Refs.\ \cite{{deol},{search}}, the resilience of
pattern-based searches to at least some of these obstacles
is confirmed in numerical experiments.)
In this 
article we add to this list the 
small, but still present, effect of a Lorenz deformation of the
circles into displaced ovals.

To derive the distortion consider 
an observer, O, who moves with speed $\beta$ relative to the CMB. At the time
the satellites measure the CMB, observer O coincides with
observer C. 

\begin{figure}[ht]
\vspace{50mm}\includegraphics{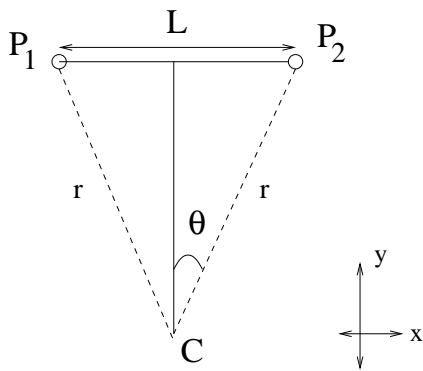} 
\vskip 5truept
\caption{Two points on opposite sides of a circle-in-the-sky from
  figure {\ref{onecirc}}. The $z$-direction is suppressed.
}
\label{C}
\end{figure}

To prove that circles deform to displaced ovals,
consider two points $P_1$ and $P_2$ on 
opposite sides of a circle of radius $L/2$ at rest with respect to C
as measured in the C frame. The two points
$P_1$ and $P_2$ lie along the axis of
motion of observer O. We take the motion to be along the $x$-axis. The
geometry for C is shown in figure \ref{C}. The emission of last scattered
photons from 
points $P_1$ and $P_2$ are two events that occur
simultaneously and a distance $L$ apart so that
$\Delta \eta  = \eta_2-\eta_1 =0 $ and $\Delta x = x_2-x_1 =L $.
The space interval, $\Delta x^\prime=x_2^\prime - x_1^\prime$, and
time interval, $\Delta \eta^\prime=\eta_2^\prime - \eta_1^\prime$, of the
events $P_1$ and $P_2$ 
as measured by the observer O are given 
by the Lorenz transformation, 
\begin{equation}
\pmatrix{ \Delta \eta^\prime \cr \Delta x^\prime}=
\pmatrix{\gamma & -\gamma \beta \cr -\gamma \beta & \gamma}
\pmatrix{\Delta \eta \cr \Delta x} \ \ ,
\label{lorenz}
\end{equation}
with $\gamma=(1-\beta^2)^{-1/2}$ so that 
\begin{eqnarray}
\Delta \eta^\prime &=& -\gamma \beta L \nonumber \\
\Delta x^\prime &=& \gamma L \ \ .
\label{expand}
\end{eqnarray}
In words, observer O does not believe that the emission of last
scattered
light occurs simultaneously but instead observes $P_2$ emit light
before $P_1$. O also observes the distance between the two points 
to be larger than in the rest frame by the factor $\gamma$. The
elongation occurs only along  
the direction of motion. The perpendicular axis is still observed to 
have length $L$ and the circle is deformed into an ellipses 
with long axis $\gamma L$ and short axis $L$.

\begin{figure}[ht]
\vspace{25mm}\includegraphics{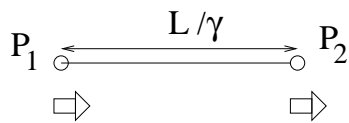} 
\vskip 5truept
\caption{O measures a Lorenz contraction of the distance between $P_1$
  and $P_2$.
}
\label{contract}
\end{figure}

Another way to derive this result is to begin with the Lorenz
contraction. Observers in the O frame would measure the intrinsic 
distance between opposite points on the circle to be Lorenz contracted to
$L/\gamma$. They would come to this conclusion by performing
the following experiment. Traveling in a rocket at speed $\beta$,
their spaceship coincides with $P_1$ at time $\bar \eta^\prime_1$
as shown in figure \ref{contract}. The
location at which this happens in their frame is the center of the
rocket
which they take to be the origin so $\bar x^\prime_1=0$. 
Some
time later the ship coincides with point $P_2$ at time $\bar
\eta^\prime_2$. Again
the location at which this happens is the center of the rocket, namely 
$\bar x^\prime_2=0$. Now, they see $P_1$ and $P_2$,
pass by their windows traveling
at speed $\beta$ and so the time elapsed $\bar \eta^\prime_2-\bar
\eta^\prime_1=\bar L^\prime/\beta$ where $\bar L^\prime$ is the length
they measure.

In C's frame, which is at rest with respect to the ring, the length
along the axis is simply $L$. C sees the rocket pass with speed
$\beta$ and determines that it reached point $P_2$ a time $L/\beta$
after it coincided with point $P_1$.
We can equate
spacetime intervals to derive the Lorenz contraction:
\begin{eqnarray}
-(\bar\eta^\prime_2-\bar\eta^\prime_1)^2 + (\bar x^\prime_2-\bar
 x^\prime_1)^2 &=& - 
 (\bar \eta_2-\bar \eta_1)+ (\bar x_2-\bar x_1)^2 \nonumber \\
-{\bar L^{\prime 2}\over \beta^2} &=& -{L^2\over \beta^2}
+L^2
\end{eqnarray}
which yields $\bar L^\prime=L/\gamma$. Observer O believes the ring is
contracted
relative to C's measure. 
However, the emission of last scattered light from the 
surface of last scattering constitutes a slightly different
measurement. In that case, O sees light emitted from point $P_2$ first
and only after the ring has continued to move for a time $|\Delta
\eta^\prime | = \gamma \beta L$ (as derived in eqn.\ (\ref{lorenz}))
does $P_1$ emit last scattered 
light.
Consequently the point $P_1$ has moved an additional distance 
$\gamma \beta^2 L$ on top of the intrinsic separation of $L/\gamma$
giving a net separation of
\begin{equation}
\gamma \beta^2 L+ L/\gamma = \gamma L
\end{equation}
and confirming the first derivation. O perceives the ring to be wider
along
the direction of motion than it is perpendicular to the direction of 
motion.

\begin{figure}[ht]
\vspace{45mm}\includegraphics{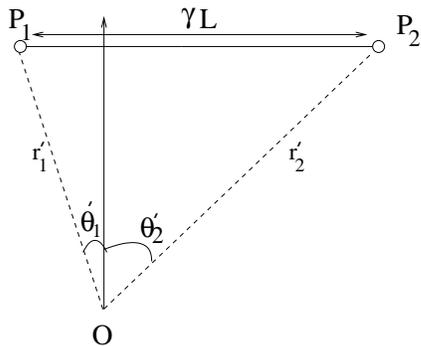} 
\vskip 5truept
\caption{The points $P_1$ and $P_2$ subtend angles $\theta_1^\prime$
  and $\theta_2^\prime$ respectively.
}
\label{O}
\end{figure}

To determine the angles subtended by the points $P_1$ and $P_2$ as
measured by O (figure \ref{O}), consider the two events, emission of
light from $P_2$ 
and the receipt of that light. In the C frame, 
light is emitted at $\eta_2$ and received at $\eta_0$ so that
$\eta_0-\eta_2=r$ where $r$ is the distance the light has
traveled. These two events are separated by $x_0-x_2=-r\sin \theta=-L/2$,
where $\theta$ is the angular radius of the circle in the stationary
frame of C.

From the Lorenz transformation, the separation of the two events
$\Delta\eta^\prime=\eta_0^\prime-\eta^\prime_2$ in time and $\Delta
x_2^\prime=x_0^\prime - x_2^\prime$ in space are given by
\begin{eqnarray}
\Delta \eta_2^\prime &=& \gamma(1+\beta\sin\theta)r=r^\prime_2 \nonumber
\\
\Delta x_2^\prime &=&
-\gamma(\sin\theta+\beta)r=-r^\prime_2\sin\theta_2^\prime
\label{th1}
\end{eqnarray}
where $r^\prime_2$ is the distance light travels in the O frame and 
$\theta_2^\prime$ is the angle subtended by point $P_2$, as measured
from 
the vertical.

Similarly for point $P_1$ it follows that
\begin{eqnarray}
\Delta \eta_1^\prime &=& \gamma(1-\beta\sin\theta)r=r^\prime_1 \nonumber
\\
\Delta x_1^\prime &=&
\gamma(\sin\theta-\beta)r=r^\prime_1\sin\theta_1^\prime \ \ ,
\end{eqnarray}
where $\theta_1^\prime$ is the magnitude of the angle $P_1$ subtends
from the verticle as drawn in figure \ref{O}.
Notice that $\Delta \eta_1^\prime-\Delta \eta_2^\prime=\eta_2^\prime -
\eta_1^\prime =-\gamma \beta L$ and 
$\Delta x_1^\prime-\Delta x_2^\prime=x_2^\prime -
x_1^\prime =\gamma L$ confirming the results of eqn.\ (\ref{lorenz}).

It follows from eqn.\ (\ref{th1}) that
\begin{equation}
\sin\theta_2^\prime={(\sin\theta+\beta)\over (1+\beta\sin\theta)}
\ \ \ .
\label{step}
\end{equation}
If we assume that $\theta_2^\prime=\theta+\delta_2$ where $\delta_2$
is small, then the angle 
can be approximated by $\sin\theta_2^\prime\simeq \sin\theta +\delta_2
\cos\theta$.
Eqn.\ (\ref{step}) then yields
\begin{equation}
\delta_2\simeq {\beta\cos\theta\over (1+\beta\sin\theta)} \ \ .
\end{equation}
The angle subtended by $P_2$ is larger than in the comoving frame
by a piece proportional to $\beta$.
Similarly, the angle subtended by $P_1$ is smaller by
$\theta_1^\prime=\theta-\delta_1$,
\begin{equation}
\delta_1\simeq {\beta\cos\theta\over (1-\beta\sin\theta)} \ \ .
\end{equation}
The center of the circle is displaced by $\delta\sim \beta$.

\begin{figure}[ht]
\vspace{25mm}\includegraphics{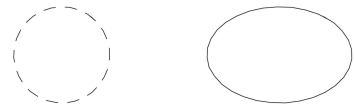} 
\caption{Observer C measures the dotted-line circle centered on $x=0$
while observer O measures the solid-line ellipse displaced by
$\gamma\beta r$.
The circle is displaced as well as distored in shape.
In the figure $r$ is taken to be $2L$ for illustration and $\beta=3/4$.
}
\label{egg}
\end{figure}

In C's frame, the circles of figure \ref{onecirc} lie in the $x-z$ plane and
are parameterized by the 4-vector
$(\eta_*-\eta_0,x-x_0,y-y_0,z-z_0)=(-r,L\cos\phi/2,0,L\sin\phi/2)$.
In O's frame, the shape of the intersection of the copies of the
surface of last scattering is parameterized by  
\begin{eqnarray}
x^\prime &=& \gamma\left ({L\cos\phi \over 2}\right )+\gamma \beta r
\nonumber \\
z^\prime &=&{L\over 2}\sin\phi \ \ 
\end{eqnarray}
and is a displaced ellipse as shown in figure \ref{egg}. Notice that the
distortion to the shape is proportional to $\gamma$ 
and so is second order in $\beta$ while the displacement is first
order in $\beta$ and is therefore the bigger effect. For circles that
are not perpendicular to the direction of motion the shape will be
slightly different from a perfect ellipse.

\begin{figure}[ht]
\vspace{50mm}\includegraphics{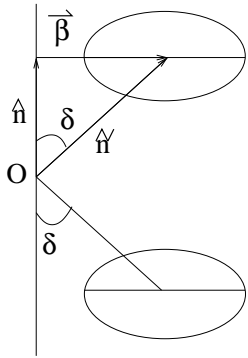} 
\vskip 5truept
\caption{In figure \protect{\ref{onecirc}} observer C sees one circle
  in front of him and one behind him. Unlike C, the Lorenz boosted
  observer O does not see one cirlce in front of him and one behind
  him.He sees the center of both ovals displaced by $\delta\sim \beta$ as
  drawn. Notice the oval pairs are opposite each other and face on but
  they are not on either side of the observer O.If he were to find an
  oval in direction $\delta$ and were to look for its pair
  in the direction $180^o+\delta$, he would miss the pair by an angle $\sim
  2\delta\sim 2\beta$ (for small $\delta$ and $\beta$).
}
\label{displace}
\end{figure}

In
the boosted frame, 
the center of one oval is
offset by $\delta$ from the vertical and its pair is not $180^o$
further on at an angle of $180^o+\delta$
but instead is centered at $180^o-\delta$. The center of the pair is
off by $2\delta$ as illustrated in figure \ref{displace}.

The above calculation applies to a specific geometry. More generally,
given any point on the surface of last scattering $(\eta_*-\eta_o,r\hat
n)=r(-1,\hat n)$, the direction to that point is Lorenz transformed to 
\begin{equation}
\hat n^\prime={\hat n+\left[(\gamma-1)\hat \beta\cdot \hat
    n+\gamma\beta\right ]\hat \beta \over \gamma (1+\vec \beta\cdot\hat
    n)} \  \ \ .
\end{equation}
(The data could be corrected by applying the inverse of this
transformation to the map, that is, if we were in uniform motion.)
In particular, if $\hat n$ is the direction to the center of a circle,
then to lowest order in $\beta$ for the case of $\hat n\cdot \vec \beta=0$, 
the center wil be displaced to
\begin{equation}
\hat n^\prime=\hat n+\vec \beta
\end{equation}
If we were to look for an antipodal pair in the opposite
direction, $-\hat n$, the pair would also be displaced to $\hat
n_{pair}^\prime\simeq -\hat n+\vec \beta=-\hat n^\prime +2\vec \beta$ which
is not $180^o$ oposite 
as illustrated in figure \ref{displace}.

The effect is largest when $\hat n\cdot \hat \beta=0$ as in figure
\ref{onecirc} and the effect vanishes when $\hat n\cdot \hat \beta=1$.
For $\beta\sim 1.23\times 10^{-3}$, the maximum angular difference in
the location of a pair from where one expects is
\begin{equation}
2\beta \sim 0.14 ^{o} \ \ .
\end{equation}
The deviation is just below WMAP's
angular resolution which at best is $<0.25^{o}$ in the 90 GHz channel
and at worst is $\sim 0.93^{o}$ in the 22 GHz channel. The 
stretching and displacement
of circles-in-the-sky would also be relevant for the future 
{\it  Planck Surveyor} which aims for an angular resolution of around
several arc-minutes.
Still, if the data is smoothed before scanned, the aberration may
be below the resolution of circle searches. It will be something to bear in 
mind for future analysis.

Distortions due to the motion of our planet and galaxy could effect
any statistical, pattern-based search, not just
circles-in-the-sky. There are other modifications to consider as well,
such as the complication that 
the earth is not in uniform motion and the extension to a curved
space.
This calculation is intended to draw out the general issue.

An intriguing possibility that is highlighted here and in earlier
articles on special relativistic effects in a finite universe
\cite{{bl1},{bl2}}, involves compactifying space{\it time} and not just
space. The universe is a (3+1)-dimensional spacetime and in principle
we should consider the geometry of this four-dimensional manifold. 
Of course, we have little idea how to go about this as much due to
difficulties of interpretation as anything else. It is a reminder that
whatever underlying principle determines the creation of the universe and
its topology will shape our future ideas on the nature
of space and time.

\section*{Acknowledgements}
I am especially grateful to Pedro Ferreira for revisiting this topic with me 
from time to time over the years. I also want to thank
John Barrow, Joe Silk, Carsten van der
Bruck, Constantinos Skordis, Glenn Starkman, Neil Cornish, and David Spergel.


\begin{thebibliography}{99}

\bibitem{cobe} G.F. Smoot et. al, Ap. J. {\bf 371} L1 (1991);
  G.F. Smoot et. al., Ap. J. {\bf 396} L1 (1992).

%C.L. Bennett et. al., Astrophys.J. 464 (1996) L1-L4.

\bibitem{wmap} C.L. Bennett et. al., submitted Ap. J.,
  astro-ph/0302207.

\bibitem{wolf} J. A. Wolf, {\it Space of Constant Curvature (5th
ed.)}, (Publish or Perish, Inc., 1994).

\bibitem{ellis} G.F.Ellis, Q.J.R.Astron. Soc. {\bf 16} 245 (1975);
G. F. R. Ellis, Gen. Rel. Grav. {\bf 2} 7(1971).

\bibitem{review1} M. Lachieze-Rey and J. -P. Luminet, Phys. Rep.
{\bf 25}, 136, (1995).

\bibitem{review2} J. Levin, Phys. Rep. 365 (2002) 251.

\bibitem{css} N.J.Cornish, D.N.Spergel and G.Starkman,
Phys. Rev. D {\bf 57} (1998) 5982.

\bibitem{spots} J. Levin, E. Scannapieco, G. Gasperis, J. Silk and
J. D. Barrow,
{\em Phys. Rev. } { D} {\bf 58} (1998) article 123006.

\bibitem{deol} A. de Oliviera-Costa, M. Tegmark, M. Zaldarriaga,
  and A. Hamilton, astro-ph/0307282. 

\bibitem{bl1} J.D. Barrow and J. Levin,Phys. Rev. A 63 (2001).

\bibitem{bl2} J.D. Barrow and J. Levin,
Mon.Not.Roy.Astron.Soc. 346 (2003) 615


\bibitem{search} N.J. Cornish, D.N. Spergel, G.D. Starkman,and
  E. Komatsu,astro-ph/0310233. 


\end{thebibliography}
\end{document}